\documentclass[
  aps,
  prd,
  reprint,
  superscriptaddress,
  longbibliography,
  nofootinbib,
  floatfix
]{revtex4-2}

\usepackage[vcentermath]{youngtab}
 \usepackage{ytableau}

\usepackage[T1]{fontenc}
\usepackage[utf8]{inputenc}
\usepackage{lmodern}
\usepackage{amsmath,amssymb,mathtools}
\usepackage{bm}
\usepackage{booktabs}
\usepackage{graphicx}
\usepackage[dvipsnames]{xcolor}
\usepackage{hyperref}
\usepackage{microtype}
\usepackage[normalem]{ulem}

\hypersetup{
 colorlinks=true,
 linkcolor=NavyBlue,
 citecolor=ForestGreen,
 urlcolor=BrickRed,
 pdftitle={$SU(\mathcal{N})$ baryon formation in the early Universe   
    and dark matter},
 pdfauthor={L. Di Luzio, S. Di Valeriano, E. Nardi}
}

\newcommand{\mR}{\mathcal R}
\newcommand{\mN}{\mathcal N}
\newcommand{\mB}{\mathcal B}
\newcommand{\mM}{\mathcal M}
\newcommand{\mC}{\mathcal C}
\newcommand{\GeV}{\,\mathrm{GeV}}

\newcommand{\mpl}{M_{\rm Pl}}
\newcommand{\Tr}{\operatorname{tr}}

\def\eq#1{{Eq.~(\ref{#1})}}

\begin{document}

\title{$SU(\mathcal{N})$ baryon formation in the early Universe   
    and dark matter}

\author{Luca Di Luzio}
\email{luca.diluzio@pd.infn.it}
\affiliation{INFN Sezione di Padova, Via Francesco Marzolo 8, 35131 Padova, Italy}

\author{Samuele Di Valeriano}
\email{sdivaler@sissa.it}
\affiliation{SISSA, International School for Advanced Studies, Via Bonomea 265, I-34136 Trieste, Italy}
\affiliation{INFN, Sezione di Trieste, via Valerio 2, 34127 Trieste, Italy}

\author{Enrico Nardi}
\email{enrico.nardi@lnf.infn.it}
\affiliation{Istituto Nazionale di Fisica Nucleare, Laboratori Nazionali di Frascati, C.P.~13, 00044 Frascati, Italy}
\affiliation{Laboratory of High Energy and Computational Physics, NICPB, R\"avala 10, 10143 Tallinn, Estonia}

\begin{abstract}

We study the formation of $\mathcal{N}$-quark baryons during the confinement crossover 
of an $SU(\mathcal{N})$ gauge theory containing quarks in the fundamental representation, 
with masses around or below the confinement scale.
We assume stepwise baryon formation through the successive fusion of pairs 
of quark-clusters, with color interaction rates obeying   
quadratic Casimir scaling.
We include the complete set of reactions between $SU(\mathcal{N})$ antisymmetric clusters 
dressed with their specific spin-flavor multiplicities.
We show that at large $\mathcal{N}$, baryon formation is hindered by a Casimir bottleneck, 
such that fusion of few-quark clusters in the initial steps of the baryon assembly 
chain is strongly outweighed by cluster destruction processes. We 
find that for $\mathcal{N}=12\,(18)$ 
the baryon-to-meson yield ratio is suppressed 
down to $10^{-12}\,(10^{-23})$ of its value at $\mathcal{N}=3$.
We embed the mechanism in a vectorlike three-flavor model whose lightest
baryon can be electrically neutral, while  
axial-anomaly-induced couplings and dimension-six
operators induce decays of the unstable mesons and charged baryons 
safely before big-bang nucleosynthesis. The relic density of cosmologically stable 
neutral $\mathcal{N}$-baryons is thus  set during confinement at values so small that, for 
sufficiently large $\mathcal{N}$, the baryon masses required to saturate 
the dark matter energy density exceed by several orders of magnitude 
the unitarity bound on annihilation cross sections.
\end{abstract}

\maketitle

\section{Introduction}
\label{sec:introduction}

Confining gauge theories offer a simple origin for cosmologically stable composite dark
matter (DM): an exact or sufficiently accurate vector charge stabilizes the
lightest baryon $\mB$, while mesons $\mM$ can decay through interactions with the
Standard Model (SM)~\cite{Antipin:2015xia,Appelquist:2015yfa,
Kribs:2016cew,Mitridate:2017oky,Asadi:2026mip}.  Relic-density calculations commonly begin
with a population of already formed baryons and determine its subsequent
$\mB$-$\bar\mB$  annihilation.  This does not need to be the appropriate initial
condition when the gauge group has many colors.  Before annihilation can act,
the plasma must first assemble a color singlet baryon from  $\mN$ constituents, while the 
competing formation of mesons from quark-antiquark pairs could, especially in the absence 
of a cosmological or local baryon asymmetry, rapidly deplete the abundance of free constituents.
For small values of $\mN$, as in QCD with $\mN=3$, the baryon-to-meson formation ratio is not subject 
to any significant suppression. The potential responsible for the formation of a diquark cluster, $\Psi+\Psi\to[\Psi^2]_A$ (where the subscript $A$ refers to 
the  antisymmetric color configuration) is  controlled by a Casimir coefficient 
smaller by only  a factor of two than that for meson formation, 
and, since $[\Psi^2]_A\in\mathbf{\bar{3}}$, the final step of $\mB$ assembly,
$\Psi+[\Psi^2]_A\to[\Psi^3]_A$, does not introduce any additional suppression.

This picture changes drastically at large 
$\mN$,\footnote{Large-$\mN$ gauge models are particularly relevant to axion physics, 
as they can protect the Peccei-Quinn symmetry  from explicit breaking by effective 
operators up to  dimension $\mN$~\cite{DiLuzio:2017tjx,Ardu:2020qmo,Lu:2023ayc,DiLuzio:2026vhj}.} 
where a barrier 
emerges in the baryon assembly chain. 
The obstruction arises already in the first reactions. The attractive potential binding a quark and an antiquark into a singlet meson is proportional to the fundamental Casimir, $C_F=O(\mN)$. By contrast, the potential attracting a quark and a small antisymmetric cluster is $O(1)$, whereas an $O(\mN)$ potential attracts an antiquark and the same cluster, resulting in a destructive process in which a meson is formed and the number of cluster constituents is decreased by one. Free quarks and bound constituents are therefore efficiently removed into mesons before most same-sign clusters can complete their assembly into $\mN$-quark baryons.
A simple and transparent description of this \emph{Casimir bottleneck}, in which only the transitions $[\Psi^p]\to[\Psi^{p\pm1}]$ between nearest neighbors (NN) in cluster space were tracked, together with its cosmological implications, was presented in the accompanying Letter~\cite{DiLuzio:2026aco}.

The purpose of this companion paper is twofold. First, we expose the group-theory and kinetic 
content of the mechanism in detail.   We go beyond the NN approximation (BNN), including  
 all fusions and rearrangements between pairs of clusters that remain in the fully antisymmetric color configuration, corresponding to the most attractive channels at each step. Channels involving 
 mixed-symmetry states are less attractive and often repulsive.
We generalise the one-flavor model presented in Ref.~\cite{DiLuzio:2026aco} to $n_f=3$, which allows for meson decays, and derive the appropriate $SU(6)$ spin--flavor weights. We compare the results for the different NN, BNN, $n_f=1$, and $n_f=3$ cases, showing that the Casimir bottleneck remains operative in all cases.
Second, we give an explicit DM realization.  
We introduce three vectorlike flavors with hypercharges $(+2/3,-1/3,-1/3)$, analogous to those of the SM light quarks. A neutral stable baryon can then arise when $\mN$ is divisible by three.  
This same flavor structure provides operators that allow the charged baryons, pseudo-Nambu--Goldstone bosons (pNGBs), and vector 
mesons to decay.

Our approach is based on controlled group-theory results combined with deliberately simple kinetic ansatzes. One-gluon exchange fixes the color coefficients for cluster-(anti)cluster interactions exactly in the Coulomb regime, while lattice calculations support approximate Casimir scaling of static potentials at intermediate 
distances~\cite{Ambjorn:1984mb,Deldar:1999vi,Bali:2000un,Mykkanen:2012ri}.  
Turning those potentials into recombination rates for relativistic light fermions necessarily 
involves an educated ansatz. Our equations for the evolution of cluster abundances are inspired by resonance recombination models (RRM)~\cite{Ravagli:2007xx,Ravagli:2008rt,He:2019vgs}, but their extrapolation to an $\mN$-step assembly chain at large $\mN$ does not correspond to a systematic RRM expansion.
We accordingly distinguish the robust hierarchy of color channels from the numerical 
yields obtained with the reference ansatzes.
Our results assume a  homogeneous confinement crossover, with no spatial inhomogeneities 
and no cosmological or local particle-antiparticle asymmetries in the $\mN$ sector. Support for these assumptions, their dependence on the number and masses of the fundamental fermions, and the 
qualitatively different dynamics of a first-order transition are discussed in Sec.~\ref{subsec:transition-order}. We also state explicitly the assumption that string breaking completes hadronization predominantly through meson
formation. These qualifications define the domain in which the output of the evolution network can 
be interpreted as the late relic abundance of $\mN$-baryons.

The paper is organized as follows.  Section~\ref{sec:model} defines the
three-flavor model and the conditions to ensure that the lightest  baryon is neutral.  
Sections~\ref{sec:casimir} and
\ref{sec:network} derive the color and spin--flavor factors and describe the
NN and BNN networks.  Section~\ref{sec:results} presents the numerical
results, and Sec.~\ref{sec:hadronization} explains their 
interpretation in terms of the hadronization model adopted.  
Section~\ref{sec:cosmology} discusses the matching between the baryon relic density 
and a DM benchmark.  App.~\ref{app:group}
collects the group-theory derivations for the mixed-symmetry channels.
App.~\ref{app:network-checks}
presents  the network equations for the NN chain. 
App.~\ref{app:decays} collects  decay formulae
for the pNGB.

\section{ $\mathbf{SU(\mN)}$  confining  sector with $\mathbf{n_f=3}$}
\label{sec:model}

\subsection{Field content and symmetries}

Consider an $SU(\mN)$ gauge theory with confinement scale $\Lambda$
and three Dirac fundamentals $\boldsymbol{\mN}$
\begin{equation}
 \Psi=(U,D,S)\,,\qquad
 U,D,S\sim\boldsymbol{\mN},
 \label{eq:flavors}
\end{equation}
 singlets of $SU(3)_c\times SU(2)_L$ and carrying vectorlike
hypercharges
\begin{equation}
 Y_U=\frac{2}{3},\qquad Y_D=Y_S=-\frac{1}{3}.
 \label{eq:hypercharges}
\end{equation}
Each Dirac pair introduces no gauge anomaly.  The renormalizable Lagrangian has an
accidental baryon-like vector symmetry $U(1)_\mB$, which is also preserved by the
dimension-six interactions retained below, under which mesons are neutral. 
Normalizing the $U(1)_\mB$ charge of the constituents to $1/\mN$, a color-singlet 
$\mN$-quark baryon carries charge $+1$, and hence the lightest baryon and antibaryon are stable.
The hypercharge of a baryon containing $ n_U+n_D+n_S=\mN$ constituents is 
\begin{equation}
 Y_{\mB}=\frac{2n_U-n_D-n_S}{3}
 =n_U-\frac{\mN}{3}.
 \label{eq:baryon-charge}
\end{equation}
Thus  when
$\mN\equiv0\;({\rm mod}\;3)$, every baryon has integer 
hypercharge, and being an $SU(2)$ singlet, 
also integer electric charge.
  There are neutral states with $n_U=\mN/3$, and in particular the three-flavor 
symmetric state with 
\begin{equation}
 n_U=n_D=n_S=\frac{\mN}{3}
 \label{eq:neutral-composition}
\end{equation}
is neutral.\footnote{More in general one can introduce $n_f$ flavors 
with hypercharges $Y_U =(n_f-1)/n_f$, $Y_{D_a}= -1/n_f$  ($a=1,\dots,n_f-1$).
For $\mN \equiv 0$ (mod$\;n_f$) every $\mN$-baryon has integer electric charge, and the baryon multiplet contains neutral states. Moreover, 
the dimension-six operators of \eq{eq:portals} are also allowed.} 
The lowest orbital state $L=0$ is symmetric, while the $\mN$-color part of the wave function is antisymmetric. Fermi statistics then requires a fully symmetric spin-flavor wave function under $SU(6)_{sf}$, with minimum allowed spin $S=0$ for even $\mN$ and $S=1/2$ for odd $\mN$.

Hypercharge self-energy can determine the neutral state as the lightest member
of the baryon multiplet.  Parametrically,
\begin{align}
 E_{\mB}={}&E_0+\sum_i n_i m_i
 +c_Y\alpha_Y  Y_{\mB}^{2}\, \Lambda+\cdots,
 \label{eq:hypercharge-self-energy}
\end{align}
with $c_Y>0$ corresponding to  the electrostatic energy of a charged composite, 
and $\alpha_Y$ to the hypercharge fine structure constant at the scale $\Lambda$. 
The neutral state is selected when the constituent mass splittings obey
\begin{equation}
 |m_i-m_j|\lesssim c_Y\alpha_Y\,\Lambda.
 \label{eq:neutral-condition}
\end{equation}
This is the analogue of the competition between quark masses and
electromagnetic self-energies in QCD hadron 
splittings~\cite{Gasser:2020mzy}.  Hyperfine corrections change the order-one boundary,
but not the mechanism.

\subsection{Decays of the heavier baryons and mesons}
\label{subsec:portals}

The 
three $\Psi^a$ flavors may couple to SM  neutral  currents,
while conversion between $U$ and $D$ or $S$ can couple to a charged
right-handed-quark current.  A representative effective Lagrangian is
\begin{align}
 \mathcal L_6={}&
 \left( \frac{a_U}{\Lambda_6^2} \bar U\gamma_\mu P_X U + 
 \ \frac{a_{ij}}{\Lambda_6^2} \bar\Psi^i\gamma_\mu P_X\Psi^j\right)
\left(\bar f_{\rm SM}\gamma^\mu f_{\rm SM}\right)
 \nonumber\\
 &+\frac{b_i}{\Lambda_6^2}
 (\bar\Psi^i\gamma_\mu P_X U)
 (\bar u_R\gamma^\mu d_R)+{\rm h.c.},
 \label{eq:portals}
\end{align}
where $P_{X}$ represents $P_L$ or $P_R$ chiral projectors, $f_{\rm SM}$ denote chiral SM fermions, and 
$i,j=D,S$. 
All operators in
Eq.~\eqref{eq:portals} preserve $U(1)_\mB$ while  allowing for heavier baryons
 to decay  to the lightest neutral one.
Upon confinement, the spontaneous breaking of the approximate
three-flavor chiral symmetry also produces a multiplet of
pNGBs. The same dimension-six operators
mediate the decays of those pNGBs that lack anomaly-induced couplings to
SM gauge bosons.
Other
neutral pNGBs decay directly through the hypercharge anomaly, while vector
mesons can decay via $SU(\mN)$ strong interactions 
to lighter pNGBs or through the currents in
Eq.~\eqref{eq:portals}.  When kinematically allowed, $SU(\mN)$  gaugeballs decay
strongly to lighter hadrons.  If the constituent mass 
is instead too high for such decays,  loops of $\Psi$'s that are charged under 
both $SU(\mN)$ and $U(1)_Y$ generate a mixed Euler--Heisenberg interaction that permits decays to
the  hypercharge (or QED) gauge boson.  The detailed pNGB classification and parametric widths are
given in App.~\ref{app:decays}.

For a baryonic cascade with available energy $\Delta E$, a useful inclusive
three-body estimate is
\begin{equation}
 \Gamma_{\mB^*}\simeq
 \frac{g_H}{8(192\pi^3)}\frac{(\Delta E)^5}{\Lambda_6^4},
 \label{eq:baryon-cascade-width}
\end{equation}
where $g_H$ counts accessible daughter channels. This corresponds to the lifetime
\begin{equation}
 \tau_{\mB^*}\simeq0.65\,{\rm s}
 \left(\frac{10^{11}\GeV}{\Delta E}\right)^5
 \left(\frac{\Lambda_6}{\mpl}\right)^4
 \left(\frac{100}{g_H}\right), 
 \label{eq:baryon-cascade-lifetime}
\end{equation}
where $\mpl=1.22\times10^{19}\GeV$ is the Planck
mass.  For  $\Lambda=10^4\GeV$ and  $\Lambda_6=10^6\GeV$
(see e.g.~the reference values adopted in 
the benchmark example in Sec.~\ref{sec:cosmology})  even splittings far below the natural $O(\alpha_Y\Lambda)$ 
scale are sufficient for decay before big-bang
nucleosynthesis (BBN).

\subsection{Order of the confinement phase transition}
\label{subsec:transition-order}

The ratio $n_\mB/n_\mM$ resulting from the $SU(\mN)$ 
phase transition (PT) depends critically on 
whether the transition is first-order or  it is rather a cross-over.
In  case the PT is first order, the baryon abundance is bounded from below 
by a peculiar effect related to bubble formation. 
The distribution of the heavy quarks and antiquarks 
trapped inside pockets of the deconfined phase~\cite{Witten:1984rs} 
is characterized by a local $\Psi$-$\bar \Psi$  asymmetry related to random  statistical fluctuations, which prevents a complete binding of all $\Psi$ and $\bar\Psi$ into mesons~\cite{Asadi:2021yml,Asadi:2021pwo,Gouttenoire:2023roe}. 
If the PT is first order, this asymmetric component  determines 
a floor for the  surviving $\mB$ and $\bar\mB$ abundances.

Lattice studies have found 
 that in pure $SU(\mN)$ Yang-Mills theories  the PT  
is always strongly first order
for $\mN\geq3$, and  becomes stronger as $\mN$
increases~\cite{Svetitsky:1982gs,Lucini:2005vg,Datta:2009jn,
Lucini:2012gg}. 
  This result holds also in the presence of very heavy quarks,  
   since in the limit  $m_\Psi \gg \Lambda$ the $SU(\mN)$ fermions decouple from the   dynamics of the PT. 
However, as the quark mass scale is lowered, the dynamical 
effects of quarks soften the transition, and 
for values of $m_\Psi$ that are still a factor of a few larger than the confining scale $\Lambda$ the PT can turn into a continuous 
 cross-over.\footnote{The  nature of 
the QCD PT between confined hadronic matter and deconfined 
quark-gluon plasma is generally visualized with the so-called Columbia plot~\cite{Brown:1990ev}.}
Ref.~\cite{Alexandrou:1998wv} studied 
the quark-mass dependence of the QCD PT, finding that with one 
flavor ($n_f=1$),  a smooth crossover behavior persists 
as long as $m_q \lesssim 10\,\Lambda_{\rm QCD}$.
Ref.~\cite{Saito:2011fs} performed a dedicated study for 
$n_f=2$ finding that in this case  the threshold until which  
the crossover behavior persists is increased 
to  $m_q \sim 20\,\Lambda_{\rm QCD}$.
Ref.~\cite{LatticeStrongDynamics:2020jwi} studied the 
PT in $SU(4)$ with four degenerate flavors, finding that  the PT remains a smooth 
cross-over as long as the quark masses remain below  few times the confining scale. 
A smooth crossover has also been observed with $n_f=2$ and $\mN=3,\,4$, and 
5~\cite{DeGrand:2021xha,DeGrand:2021zjw} (see also~\cite{Ayyar:2018ppa,DeGrand:2018tzn}). However, to our knowledge, there are no lattice data for larger values of $\mN$. Thus,  whether a homogeneous crossover remains plausible for $\mN=\mathcal{O}(10)$ with $n_f=3$ remains speculative. Throughout this paper, we adopt a smooth crossover as the baseline cosmological history in order to highlight the Casimir bottleneck effect. Nonetheless, even if the transition is non-homogeneous and local statistical asymmetries become important, the hadronization of the symmetric quark-antiquark component will still be affected by the bottleneck effect.

\section{Color dynamics and the Casimir bottleneck}
\label{sec:casimir}

\subsection{Casimir coefficients and scaling}
\label{sec:Casimirscaling}

For two color sources $a$ and $b$, projection onto a definite two-body
 representation $[ab]$ gives the channel coefficient 
\begin{equation}
 \mC_{ab}=\frac12\left(C_{[ab]}-C_a-C_b\right),
 \label{eq:two-body-Casimir}
\end{equation}
where $C_a$, $C_b$, and $C_{[ab]}$ are the quadratic Casimirs of the corresponding representations.  
The  potential between two static color sources 
in a given  channel is well described by 
the Cornell form~\cite{Eichten:1974af,Eichten:1978tg,
Eichten:1979ms}
\begin{equation}
 V_{ab}(r)=\hat\mC_{ab}
 \left(\frac{\alpha}{r}-\sigma r\right),
 \label{eq:perturbative-potential}
\end{equation}
where $\alpha$ and $\sigma$ are the (positive) Coulomb and string-tension
coefficients of the reference singlet-meson potential, and
$\hat\mC_{ab}=\mC_{ab}/C_F$,
where $C_F=(\mN^2-1)/(2\mN)$ is the quadratic Casimir of the fundamental.
Thus, according to \eq{eq:two-body-Casimir},
the coefficient of the potential for a quark and an antiquark in the singlet meson channel $\mM \sim [\Psi\bar\Psi] $ 
is obtained from    
$\mC_{\Psi\bar\Psi} = \frac{1}{2}\left(0-C_F-C_F\right)=-C_F$, corresponding 
to  
$\hat\mC_{\Psi\bar\Psi}=-1$.  At short distance,
$\alpha=C_F g_s^2/(4\pi)$, 
where $g_s$ is the $SU(\mN)$ strong coupling constant,
and Eq.~\eqref{eq:two-body-Casimir} follows exactly
from one-gauge-boson exchange.  
Negative coefficients
correspond to attractive
channels that can  lead to bound state formation, 
while positive coefficients 
correspond to repulsive 
channels.
  Quadratic-Casimir scaling is therefore a perturbatively motivated starting point, whose validity 
in the intermediate-distance regime is supported 
by lattice studies~\cite{Ambjorn:1984mb,Deldar:1999vi,Bali:2000un,Mykkanen:2012ri}.
The normalization adopted for the coefficients $\hat\mC_{ab}$ makes the large-$\mN$ counting transparent.  In the
't~Hooft limit $g_s^2\mN$ is held fixed and $C_F\sim\mN$, so the meson-channel reference 
coefficient $\alpha=C_Fg_s^2/(4\pi)$ 
is $O(\mN^0)$.
We likewise define $\sigma$
from the meson channel.  The common quantities $\alpha$, $\sigma$, and the
overall confinement scale are therefore held fixed when comparing groups,
so the $\mN$ dependence of the potential in Eq.~\eqref{eq:perturbative-potential}
resides in the normalized channel factors $\hat\mC_{ab}$.  
Clearly, the structure of the baryon assembly chains also depends on $\mN$: recombination and destruction reactions span the cluster ranks $p=1,\ldots,\mN$ and involve $\mN$-dependent spin-flavor projections.

Nonperturbatively, approximate Casimir scaling of static potentials 
is observed over the intermediate-distance range
before screening and string breaking become 
dominant~\cite{Ambjorn:1984mb,Deldar:1999vi,Bali:2000un,Mykkanen:2012ri}.  
In a charge-symmetric potential, charge-conjugation-odd invariant tensors, and in particular cubic ones, are forbidden by symmetry.
  The first allowed violations therefore involve
quartic and higher even invariants and first occur at three loops~\cite{Anzai:2010td,Shevchenko:2001ij}.  
These facts make the quadratic Casimir  hierarchy a 
well-motivated leading organization of quark clusters
interactions at short and intermediate distances.  
At larger distances, string breaking and pair creation produce effects that are not captured by the network of Casimir-controlled reactions, and are eventually responsible for the complete screening of free color charges. However, as discussed in Sec.~\ref{sec:hadronization}, they cannot reduce the abundance of baryons already formed and are unlikely to appreciably increase it.

\subsection{Antisymmetric clusters}

\begin{figure}[t]
 \centering
 \includegraphics[width=0.98\columnwidth]{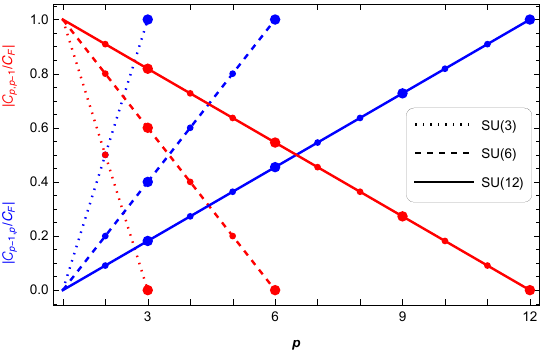}
 \caption{Magnitude of the Casimir coefficients, normalized to $C_F$, for
 cluster formation (blue) and cluster destruction
 (red).  Dotted, dashed, and solid curves correspond to
 $SU(3)$, $SU(6)$, and $SU(12)$, respectively.}
 \label{fig:casimir}
\end{figure}

 To represent clusters of $p$-quarks 
in the color antisymmetric configuration 
we introduce the following notation
\begin{equation}
 Q_p\equiv[\Psi^p]_A,\qquad
 \bar Q_p\equiv[\bar\Psi^p]_A,\qquad p=1,\ldots,\mN,
 \label{eq:cluster-definition}
\end{equation}
so that $Q_{\mN}=\mB$ is a color-singlet baryon.
The quadratic Casimir for $Q_p$ and 
 $\bar Q_p$ is
\begin{equation}
 C_p 
 =\frac{p(\mN-p)(\mN+1)}{2\mN},
 \quad C_p=C_{\mN-p}.
 \label{eq:antisymmetric-Casimir}
\end{equation}
From equation~\eqref{eq:two-body-Casimir} we  obtain  the coefficients for the 
same-sign cluster fusion 
$Q_p+Q_q \to Q_{p+q}$ ($p+q\leq \mN$) and 
cluster-anticluster rearrangement 
$Q_p+\bar Q_q \to Q_{p-q} + [Q\bar Q]^q$ ($p \geq q$) with $[Q\bar Q]^q\sim \mM^q$
\begin{equation}
 \mC^F_{pq}
 =-\frac{\mN+1}{2\mN}pq,
\qquad 
  \mC^R_{p\bar q} =-\frac{\mN+1}{2\mN} 
  (\mN-p)q\,,
 \label{eq:rearrangement-Casimir}
\end{equation}
where in the second expression $p\leftrightarrow q$ if $p<q$.
Diquark formation $Q_1+ Q_1 \to Q_2$ has
\begin{equation}
    \mC^F_{11}=-\frac{\mN+1}{2\mN},
\end{equation}
whereas for 
 direct meson formation $Q_1+\bar Q_1 \to \mM$
\begin{equation}
 \mC^R_{1\bar1}=-C_F
 =-\frac{\mN^2-1}{2\mN}\,.
 \label{eq:meson-Casimir}
\end{equation}
Their ratio is
\begin{equation}
\frac{\mC^F_{11}}
 {\mC^R_{1\bar1}}=\frac{1}{\mN-1}, 
 \label{eq:first-hierarchy}
\end{equation}
and, for the tower of NN 
channels studied in~\cite{DiLuzio:2026aco},
\begin{equation}
\frac{\mC^F_{p1}}{\mC^R_{p\bar 1}} =\frac{p}{\mN-p}\,. 
\end{equation}
Thus, for large $\mN$, the early stages of cluster formation are strongly inhibited, as destruction interactions are much stronger than the corresponding formation processes.
Figure~\ref{fig:casimir} displays the  magnitudes of the normalized coefficients for fundamental
fusion and destruction channels.  
The two lines cross at $p\simeq\mN/2$, after which the Casimir coefficients for the formation of large clusters exceed those for the corresponding destruction channels.
However, as explained  in~\cite{DiLuzio:2026aco}, 
populating this  region of cluster space  
is difficult, because meson formation 
via  direct and destruction channels rapidly depletes the abundance of free constituents.

\subsection{Rates for cluster formation}

We model the thermally averaged formation cross section in a given channel $r$ via the ansatz
\begin{equation}
 \langle\sigma v\rangle_r
 =\frac{1}{\Lambda^2}
 |\hat\mC_r|^2 P_r, 
 \label{eq:rate-ansatz}
\end{equation}
where  $P_r$, to be specified below,  denotes the spin-flavor factor for the reaction $r$.
The dependence of the rates on the square of the Casimir coefficients is motivated by the 
Coulombic part of the potential, for which the color dependence of the amplitude 
is proportional, in the one-gluon-exchange approximation,  to the Casimir coefficient. 
However, for the intermediate distance regime,  Casimir scaling by itself does not fix this exponent.  The
results should therefore be read as the consequence of a Casimir-squared two-body recombination model, with the analytic hierarchy in Eq.~\eqref{eq:first-hierarchy} being the robust theoretical input.
The factor $\Lambda^{-2}$ is introduced on dimensional grounds but drops out when comparing the rates for different processes, and it is convenient to absorb it into the definition of a dimensionless time variable.
Denoting by $\tau$ the time
elapsed from the onset of the homogeneous formation stage, we follow the Letter~\cite{DiLuzio:2026aco} and define 
\begin{equation}
 u\equiv\Lambda\tau.
 \label{eq:dimensionless-time}
\end{equation}
Note that normalizing to $C_F$ is  essential 
not only to factor out the $\mN$ dependence into 
the $\hat\mC_{r}$ coefficient, but also for comparing different $\mN$: using the unnormalized  $\mC_{r}$  merely rescales all rates by $C_F^2$, so the same trajectory for  cluster abundances  is recovered after the  rescaling  $u=C_F^2 u_{\rm unnorm}$.

\section{Reaction networks and spin-flavor factors}
\label{sec:network}

\subsection{Symmetric $SU(6)$ multiplicities}

For a ground state $p$-cluster with  $L=0$, 
and for $n_f$ flavors, the dimension of the symmetric 
spin-flavor representation of $SU(2n_f)$ 
is equal to the number of ways of distributing $p$  identical quarks among $2n_f$ spin-flavor states:
\begin{equation}
 d_p^{(2n_f)}
 =\binom{p+2n_f-1}{p} =\binom{p+2n_f-1}{2n_f-1}.
 \label{eq:sf-dimension}
\end{equation}
The completely symmetric representation of rank $p+q$ appears once in the
tensor product of the rank-$p$ and rank-$q$ symmetric representations.
Assuming uniformly populated initial states and spin-flavor
blind dynamics, the inclusive fusion probability is therefore
\begin{equation}
 s_{2n_f}(p,q)=
 \frac{d_{p+q}^{(2n_f)}}{d_p^{(2n_f)}d_q^{(2n_f)}}.
 \label{eq:sf-factor-general}
\end{equation}
For $n_f=1$  only the $SU(2)$ spin symmetry is relevant. The corresponding factors for the BNN and NN networks are, respectively,
\begin{equation}
 s_2(p,q)=\frac{p+q+1}{(p+1)(q+1)},
 \quad\  s_2(1,p)=\frac{p+2}{2(p+1)},
 \label{eq:su2-factor}
\end{equation}
whereas for $n_f=3$ the spin-flavor symmetry is $SU(6)$, which gives
\begin{equation}
 s_6(p,q)=
 \frac{\binom{p+q+5}{5}}
 {\binom{p+5}{5}\binom{q+5}{5}},
 \qquad s_6(1,p)=\frac{p+6}{6(p+1)}, 
 \label{eq:su6-factor}
\end{equation}
For quark--antiquark recombination, 
under  $SU(2)_{\rm s}\times SU(3)_{\rm f}$ 
 the tensor product of the fundamental and antifundamental representations of $SU(6)$ branches as
\begin{equation}
 \boldsymbol6\otimes\overline{\boldsymbol6}
 =\boldsymbol1\oplus\boldsymbol{35}
 \to(\boldsymbol1,\boldsymbol1)\oplus(\boldsymbol3,\boldsymbol1)
 \oplus(\boldsymbol1,\boldsymbol8)\oplus(\boldsymbol3,\boldsymbol8)
 \label{eq:meson-branching}
\end{equation}
The dimensions sum to
$36$, so inclusive pseudoscalar (9 states) and vector meson (27 states) formation has unit
spin--flavor weight.  We also take an inclusive unit factor for
cluster--anticluster rearrangement.  The latter is an explicit rearrangement 
prescription that assumes multimeson formation. All in all, 
the spin-flavor factors in \eq{eq:rate-ansatz} are
\begin{equation}
    P^F_{pq} = s_{(2 n_f)} (p,q), \qquad   P^R_{p\bar q} =1\,.
    \label{eq:PFRpq}
\end{equation}
for  fusion and rearrangement reactions, respectively.

\subsection{The NN  and BNN networks}

The NN network described in Ref.~\cite{DiLuzio:2026aco}
 considers only reactions between  a given cluster and 
a single constituent:
\begin{align}
 Q_1+Q_p&\longrightarrow Q_{p+1},
 \label{eq:nn-growth}\\
 \bar Q_1+Q_p&\longrightarrow Q_{p-1}+\mM,
 \label{eq:nn-destruction}
\end{align}
where $p=1,\ldots,\mN-1$, and 
the charge-conjugate reactions are  understood.  
Denoting by  $Q_0$  the absence of clusters, 
$p=1$ in Eq.~\eqref{eq:nn-destruction} corresponds to direct meson formation.
A virtue of the NN network is that all additional
mixed-symmetry channels are repulsive, so from the point 
of view of bound state formation  the reaction network is complete, see Appendix~\ref{app:group}.

The BNN network extends the set of reactions by including 
 all channels between totally  antisymmetric clusters, corresponding to the maximally attractive 
cluster-(anti)cluster reactions
\begin{align}
 Q_p+Q_q&\longrightarrow Q_{p+q},
 &&p+q\leq\mN,
 \label{eq:bnn-fusion}\\
 Q_p+\bar Q_q&\longrightarrow  Q_{p-q}+q\mM,
  && p\geq q,
 \label{eq:bnn-rearrangement}
\end{align}
where 
 the charge-conjugate of the same-sign fusions in \eq{eq:bnn-fusion} are
understood,  
and in the right-hand side (RHS)  of the second relation  
$p\leftrightarrow q$ if $p<q$. 
Note that the BNN network is complete only within the modeled antisymmetric $L=0$ tower. It does not include, among other effects,  mixed-symmetry channels, which are less attractive and often repulsive,  
see App.~\ref{app:group}; excited orbital states with $L>0$, which are likely to be strongly suppressed by kinematic effects; or string-breaking effects, which can become relevant particularly  at later stages, when the low number densities of color charges push the dynamics into the long-distance regime.

\begin{table}[t]
 \centering
 \renewcommand{\arraystretch}{1.32}
 \begin{tabular}{c@{\quad}c@{\quad}c@{\qquad}c@{\quad}c}
  \toprule
  &  \multicolumn{2}{c}{$n_f=1\qquad$} & \multicolumn{2}{c}{$n_f=3\ $} \\
  $\mN$
  & $y_{\mB}^{\rm NN}$ 
  & $y_{\mB}^{\rm BNN}$ 
  & $y_{\mB}^{\rm NN}$ 
  & $y_{\mB}^{\rm BNN}$ \\
  \midrule
  $3$  & $5.1\times10^{-2}$  & $4.5\times10^{-2}$
       & $3.5\times10^{-2}$  & $3.0\times10^{-2}$\\
  $6$  & $1.5\times10^{-4}$  & $2.7\times10^{-4}$
       & $2.2\times10^{-5}$  & $4.3\times10^{-5}$\\
  $9$  & $5.0\times10^{-8}$  & $2.8\times10^{-7}$
       & $8.7\times10^{-10}$ & $4.9\times10^{-9}$\\
  $12$ & $4.3\times10^{-12}$ & $5.3\times10^{-11}$
       & $6.7\times10^{-15}$ & $7.8\times10^{-14}$\\
  $15$ & $1.5\times10^{-16}$ & $3.1\times10^{-15}$
       & $1.8\times10^{-20}$ & $3.4\times10^{-19}$\\
  $18$ & $2.5\times10^{-21}$ & $7.5\times10^{-20}$
       & $2.0\times10^{-26}$ & $5.6\times10^{-25}$\\
  $21$ & $2.3\times10^{-26}$ & $9.5\times10^{-25}$
       & $1.2\times10^{-32}$ & $4.5\times10^{-31}$\\
  \bottomrule
 \end{tabular}
  \caption{Baryon pseudo-yields at the common dimensionless time
 $u=200$.  NN retains only Eqs.~\eqref{eq:nn-growth} and
 \eqref{eq:nn-destruction}; BNN includes all reactions in
 Eqs.~\eqref{eq:bnn-fusion} and \eqref{eq:bnn-rearrangement}.}
  \label{tab:network-yields}
\end{table}

Following Ref.~\cite{DiLuzio:2026aco}, let us define the 
pseudo-yield for a given $j$-cluster as 
$y_j\equiv n_j/\Lambda^3$, where $n_j$ is the corresponding number density, 
and a dot denoting $d/du$. In the BNN network, the equation for the evolution of $y_j$ 
can be written in the compact form
\begin{equation}
    \dot y_j = \sum_r (b_j^r-a_j^r)\, \mR_r\,.
    \label{eq:dotyj}
\end{equation}
In this equation, $b_j^r$ ($a_j^r$) is the number of $j$-clusters 
produced (destroyed) in  reaction $r$. 
The two types of reactions, fusion and rearrangement,  
can be written as:
\begin{align}
\label{eq:RFpq}
   & \mR^F_{pq} = \frac{1}{1+\delta_{pq}}\, y_p\, y_q \, |\hat \mC^F_{pq}|^2 \, P^F_{pq},\\
   \label{eq:RRpq}
   & \mR^R_{p\bar q} =  y_p\, y_{\bar q} \, |\hat \mC^R_{p\bar q}|^2 \,.
\end{align}
The first factor on the RHS of \eq{eq:RFpq} accounts for the symmetry 
factor of $1/2$ required for $p=q$ to avoid double counting the $Q_p+Q_p\to Q_{2p}$ reaction. The spin-flavor factor $P^F_{pq}$ is given in \eq{eq:PFRpq}, while in the second equation we have omitted $P^R_{p\bar q}=1$ for simplicity.
Thus, according to \eq{eq:dotyj} a fusion $Q_p+Q_q\to Q_{p+q}$  
contributes $-\mathcal R^F_{pq}$ to the incoming abundances $y_p$ and $y_q$,
and $+\mathcal R^F_{pq}$ to $\dot y_{p+q}$.  
The charge-symmetric initial condition is
\begin{equation}
\hspace{-1mm}
 y_1(0)=\bar y_1(0)=1,
 \   y_{p>1}(0)=\bar y_{p>1}(0)=y_{\mM}(0)=0.
 \label{eq:initial-condition}
\end{equation}
Here $y_1$ is summed over all internal spin--flavor states.  
For all $\mN$, the networks are integrated up to the common dimensionless time $u=200$. This is simply a reference stopping prescription, ensuring that, for every $\mN$, the pseudo-yields of all clusters have reached their asymptotic values.
The reaction network  conserves exactly the symmetric condition 
\begin{equation}
 \sum_{p=1}^{\mN}p(y_p-\bar y_p) =0 
 \label{eq:dark-charge-conservation}
\end{equation}
and, since pair creation from the vacuum is not included, 
before mesons are removed by decays the total constituent
number is also conserved:
\begin{equation}
 \sum_{p=1}^{\mN}p(y_p+\bar y_p)+2y_{\mM}
 = \mathrm{const}.
 \label{eq:constituent-conservation}
\end{equation}
These identities provide useful consistency checks 
for the integration code.

\begin{figure}[t]
 \centering
 \includegraphics[width=\columnwidth, height=6cm]{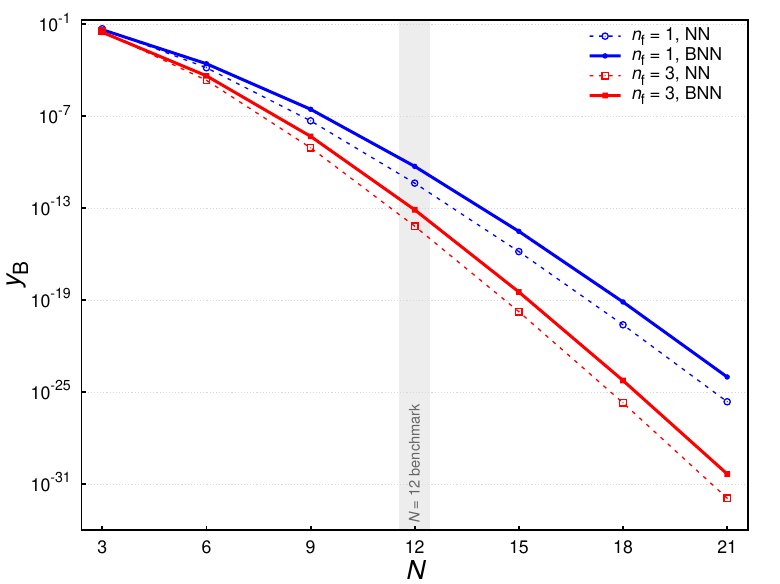}
 \vspace{-6mm}
 \caption{
 Dependence of the one-sector baryon pseudo-yield on  $\mN$ at $u=200$. Blue curves correspond to $n_f=1$, red curves to  $n_f=3$. Dashed and solid curves denote the NN and BNN  networks, respectively. The gray band marks the $\mN=12$ DM benchmark of
 Sec.~\ref{sec:cosmology}. 
 Points are computed values. Connecting lines are included only to  guide the eye.}
 \label{fig:yield-scaling}
\end{figure}

\section{Numerical effect of the Casimir bottleneck}
\label{sec:results}

We collect in Table~\ref{tab:network-yields}  the
one-sector baryon pseudo-yield $y_{\mB}=y_{\mN}$ at $u=200$.  Charge symmetry
implies $\bar y_{\mB}=y_{\mB}$.  The four columns differ only in the set of reactions, NN or BNN, and in whether cluster fusion is weighted by the 
one-flavor $SU(2)$ spin factor or the three-flavor $SU(6)$ spin-flavor factor.
At $\mN=3$, BNN gives a slightly smaller result than NN. This is because the only new
topology is $Q_2+\bar Q_2\to2\mM$, which removes diquarks that could otherwise
capture a third quark and produce  baryons, see also App.~\ref{app:network-checks}. For $\mN\geq6$, same-sign cluster
fusions bypass several sequential single-quark captures and instead enhance
the yield. In the three-flavor network, the BNN-to-NN enhancement grows from
approximately $2.0$ at $\mN=6$ to $39$ at $\mN=21$. 
Thus, the additional reactions included in the BNN network do not overcome the bottleneck, confirming that the NN approximation remains qualitatively reasonable.
In contrast, the cumulative effect of the spin-flavor  factors in suppressing the baryon pseudo-yield is quite important. For both the NN and BNN networks, the $SU(6)$ result is smaller than its one-flavor counterpart by factors ranging approximately from $10^3$ to $10^6$ for the larger values of $\mN$ between 12 and 21. Consequently, the  three-flavor BNN baryon pseudo-yield is suppressed by almost 29 orders of magnitude between $\mN=3$ and $\mN=21$.

Figure~\ref{fig:yield-scaling} shows that the suppression with $\mN$ is much
steeper than the difference between the 
 NN and BNN reaction sets. For
$\mN\geq6$, BNN multiquarks cluster-(anti)cluster 
reactions enhance baryon formation, reaching a factor of
about $38$ for three flavors at $\mN=21$. However,  they do not change the qualitative large-$\mN$ suppression.

\begin{figure}[t]
 \centering
 \includegraphics[width=0.98\columnwidth]
 {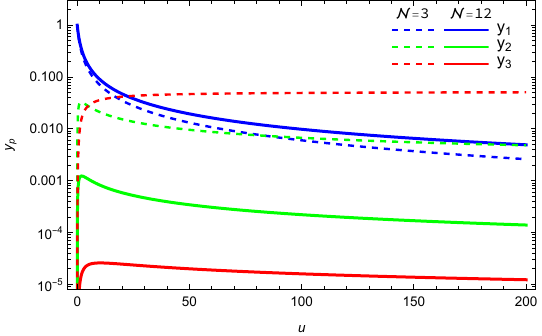}
 \caption{Evolution of the normalized abundances of free quarks (blue),
 diquarks (green), and three-quark clusters (red) in the one-flavor NN
 network.  Dashed curves show $SU(3)$ and solid curves show $SU(12)$.
 All Casimir coefficients are normalized to $C_F$, so the horizontal axis is
 the common dimensionless time of Eq.~\eqref{eq:dimensionless-time}.}
 \label{fig:evolution}
\end{figure}

Following the time evolution of small clusters 
pseudo-yields makes clear how the suppression arises.  
Figure~\ref{fig:evolution} shows the NN trajectories for the pseudo-yields of the first three clusters, $Q_1,Q_2,Q_3$, in $SU(3)$ and $SU(12)$ for $n_f=1$.
In $SU(3)$, $Q_3=\mB$, and we see that the baryon pseudo-yield rapidly reaches the percent level. In contrast, in $SU(12)$, the $Q_3$ pseudo-yield is suppressed to values of $O(10^{-5})$. Each subsequent step toward the formation of larger clusters faces further growth-versus-destruction competition. In the meantime, free quarks are drained into mesons.

\section{Completion of hadronization}
\label{sec:hadronization}

The cluster networks describe the Casimir-dominated stage, but cannot 
drive the system to complete hadronization because, when the densities 
of free color charges drop to sufficiently low values, the network 
reactions naturally freeze out.
Defining the constituent-weighted abundance of all nonsinglet clusters as
$y_{\rm col}=\sum_{p=1}^{\mN-1}p \, y_p$, we find that for the BNN runs with $\mN=12$--$21$, a small colored population remains at $u=200$:
$y_{\rm col}\simeq5\times10^{-3}$, largely independent of $\mN$. This population is strongly dominated by small $p$-clusters, mainly free quarks and, to a lesser extent, diquarks.

This is due to the fact that our approach does not include the long-distance dynamics dominated by color-string fragmentation, which is ultimately responsible for the complete hadronization of the residual colored-cluster abundance into mesons and baryons. Once the separation between naked color charges reaches the string-breaking length, pair creation fragments the color flux. At these distances, the $\mN$-ality of the cluster configuration plays a crucial role. Since the vacuum has zero $\mN$-ality, only configurations with zero total $\mN$-ality can be completely screened by pair creation into color singlets.

Thus, a pair of quarks and antiquarks can readily hadronize into two mesons,
$Q_1+\bar Q_1\to2\mM$, a process weighted by the square of the residual one-cluster abundance $y_1^2$. By contrast, to produce a baryon through the creation of a single quark-antiquark pair from the vacuum, the only initial configuration with zero $\mN$-ality is $Q_{\mN-1}+Q_1\to\mB+\mM$, whose rate is weighted by the product of  pseudo-yields $y_1y_{\mN-1}$.
In the output of our networks, this product remains at most at the level of a few percent of $y_\mB$, 
so that  this late hadronization channel cannot 
significantly alter our results.
Screening other zero $\mN$-ality configurations, such as $Q_{\mN-q}+Q_q$, requires the creation of $q$ quark-antiquark pairs and, besides the $y_q y_{\mN-q}$ suppression, is also exponentially suppressed by the associated energy cost. This qualitative argument supports a meson-dominated completion of hadronization. A dedicated lattice
hadronization calculation would be needed to confirm this conclusion.

\section{Dark matter benchmark}
\label{sec:cosmology}

The  normalization $y_1(0)=\bar y_1(0)=1$ 
of the pseudo-yields  corresponds to a density of elementary constituents  of order $\Lambda^3$ at 
a crossover temperature $T_c\simeq\Lambda$.  
Given that almost all the quarks hadronize into 
mesons, we also have $n_\mM \sim \Lambda^3$,
so that the final baryon pseudo-yield    
$y_\mB$ conveniently provides  an estimate of 
the ratio $n_\mB/n_\mM$. 
The conversion to cosmological baryon and antibaryon yields, with the usual normalization to the entropy density, is given by 
\begin{equation}
 Y_{\mB+\bar\mB}
 =\frac{45}{2\pi^2g_{*s}(T_c)}\, 2 y_{\mB}, 
 \label{eq:pseudoyield-conversion}
\end{equation}
where we took $T_c = \Lambda$ and 
$2 y_{\mB}$ accounts for the inclusion  of  the charge-conjugate sector.  We take
$g_{*s}=100$ in the numerical examples; the confined strong-sector states
are not counted as relativistic degrees of freedom after the crossover.

The present fractional contribution to the 
cosmological energy density is
\begin{equation}
 \Omega_{\mB+\bar\mB}h^2
 =2.74\times10^8
 \left(\frac{m_{\mB}}{\mathrm{GeV}}\right)
 Y_{\mB+\bar\mB},
 \label{eq:omega}
\end{equation}
where we used $\Omega_{\mB+\bar \mB}=m_{\mB}Y_{\mB}s_0/\rho_c$, with $s_0$ and $\rho_c$ the present
entropy and critical density, respectively, and
$s_0h^2/\rho_c=2.74\times10^8~\mathrm{GeV}^{-1}$.
Using $\Omega_{\rm DM}h^2=0.12$~\cite{Aghanim:2018eyx} and the three-flavor BNN baryon pseudo-yields, one can obtain the baryon mass required to account for the entirety of the DM abundance.
The values obtained for the different $\mN$ are 
listed in Table~\ref{tab:dm-matching}.
\begin{table}[t]
 \centering
 \vspace{3mm}
 \renewcommand{\arraystretch}{1.12}
 \begin{tabular}{c@{\qquad}c@{\qquad}c}
  \toprule
  $\mN$ & $Y_{\mB+\bar\mB}$ & $m_{\mB} (\Omega_\mB = \Omega_{\rm DM})$\\
  \midrule
  $12$ & $3.5\times10^{-15}$ & $1.2\times10^5\ \GeV$\\
  $15$ & $1.5\times10^{-20}$ & $2.9\times10^{10}\ \GeV$\\
  $18$ & $2.6\times10^{-26}$ & $1.7\times10^{16}\ \GeV$\\
  $21$ & $2.1\times10^{-32}$ & $2.1\times10^{22}\ \GeV$\\
  \bottomrule
   \end{tabular}
   \caption{Relic matching for the three-flavor BNN network, with
 $g_{*s}=100$, no late entropy injection, and no post-confinement
 baryon--antibaryon depletion.}
  \label{tab:dm-matching}
\end{table}
A late entropy release, for example, from a transient period of matter
domination driven by non-relativistic mesons before their decay, would
further dilute the baryon abundance and therefore increase the baryon mass
required to reproduce the observed DM density.  
Baryon annihilation after confinement is expected to be exponentially suppressed with $\mN$~\cite{Witten:1979kh} and thus to have negligible effects. In any case, any subsequent annihilation can only further decrease the abundance.
For a symmetric population,
$Y_{\mB}=Y_{\bar\mB}$, and neglecting the equilibrium density after
confinement, the Boltzmann equation during an adiabatic,
radiation-dominated epoch reads
\begin{equation}
 \frac{dY_{\mB}}{dx}
 =
 -\frac{s\langle\sigma_{\rm ann}v\rangle}{Hx}\,
 Y_{\mB}^{\,2},
 \qquad
 x\equiv\frac{m_{\mB}}{T},
 \label{eq:postformation-boltzmann}
\end{equation}
where $s$ is the entropy density. 
Integrating from the formation epoch to late times gives
\begin{equation}
 \frac{1}{Y_{\mB}^{\infty}}
 =
 \frac{1}{Y_{\mB}^{\rm form}}
 +
 \int_{x_{\rm form}}^\infty
 dx\,\frac{s\langle\sigma_{\rm ann}v\rangle}{Hx}.
 \label{eq:integrated-annihilation}
\end{equation}
Defining the  yield eventually surviving annihilation by
\begin{equation}
 \frac{1}{Y_{\mB}^{\rm ann}}
 \equiv
 \int_{x_{\rm form}}^\infty
 dx\,\frac{s\langle\sigma_{\rm ann}v\rangle}{Hx},
\end{equation}
the final abundance takes the compact form
\begin{equation}
 \frac{1}{Y_{\mB}^{\infty}}
 =
 \frac{1}{Y_{\mB}^{\rm form}}
 +
 \frac{1}{Y_{\mB}^{\rm ann}}.
 \label{eq:harmonic-yield}
\end{equation}
The Casimir-network prediction for baryon formation is therefore preserved
in the production-limited regime
$Y_{\mB}^{\rm form}\ll Y_{\mB}^{\rm ann}$, for which
$Y_{\mB}^{\infty}\simeq Y_{\mB}^{\rm form}$. 

A quantitative determination of $Y_{\mB}^{\rm ann}$ remains
model dependent.  Classic large-$\mN$ arguments suggest an exponential
suppression of exclusive baryon-pair production from a fixed number of
mesons, but they do not determine a precise exponential law for the
inclusive $\mB\bar\mB$ annihilation rate.  In particular, annihilation can
populate final states containing $O(\mN)$ mesons, whose multiplicities and
phase space are not fixed by large-$\mN$ counting alone
\cite{Witten:1979kh}. 
A qualitative comparison of the yields from  formation and  freeze-out 
of annihilation was presented in Ref.~\cite{DiLuzio:2026aco}, showing that, 
for large $\mN$ and for phenomenologically interesting values of $\Lambda$,  
the former determines the relic abundance.

Let us now present an explicit numerical example 
for the case $\mN=12$.  We assume the 
following values: 
\begin{equation}
 \begin{split}
 \Lambda& =10^4\GeV, \qquad 
 m_U\simeq m_D\simeq m_S=0.1\,\Lambda,\\
 \Lambda_6&=10^6\GeV,
 \qquad m_{\mB}\simeq\mN\Lambda=1.2\times10^5\GeV.
 \end{split}
 \label{eq:N12-benchmark}
\end{equation}
The neutral baryonic bound state is $U^4D^4S^4$ and can have spin $J=0$.  Splittings
$|m_i-m_j|\lesssim10^{-2}\Lambda$ are compatible with
Eq.~\eqref{eq:neutral-condition} and leave all three flavors equally populated at
confinement. The absence of spin and electric charge eliminates the leading magnetic-dipole interaction. The remaining electromagnetic form factors are suppressed by the compositeness scale, making $\mB$ a promising DM candidate. 
Prospects for possible detection techniques are briefly 
discussed in the next section.

The  confinement dynamics also determines the
pNGB parameters. For an order-one chiral coefficient, 
naive dimensional analysis gives
\begin{align}
 f_\pi&\simeq\frac{\sqrt{\mN}}{4\pi}\Lambda
 \simeq2.8\times10^3\GeV,
 \nonumber\\
 m_\pi&\simeq\sqrt{2m_\Psi\Lambda}
 \simeq4.5\times10^3\GeV.
 \label{eq:N12-pion-scales}
\end{align}
Here $m_\Psi$ denotes the common constituent mass in the nearly degenerate
three-flavor limit.
The pNGB decay modes are summarized in
Appendix~\ref{app:decays}.
The non-anomalous pNGBs decays proceed  through the dimension-six flavor currents in
Eq.~\eqref{eq:portals}.  When a top-quark channel is kinematically open, the
helicity suppression in Eq.~\eqref{eq:direct-pngb-width} is mild, giving the
representative lifetime
\begin{equation}
 \tau_{\pi}^{\rm dir}\sim5\times10^{-15}\,{\rm s}
 \left(\frac{\Lambda_6}{10^6\GeV}\right)^4,
 \label{eq:N12-pion-lifetime}
\end{equation}
up to order-one chiral and Wilson coefficients.  The anomalous neutral states
$\pi_8^0$ and $\eta'_D$ instead decay through the hypercharge anomaly.  
The corresponding widths are given in Eqs.~\eqref{eq:pi8-width} and \eqref{eq:eta-width}. The heavier baryon-multiplet states are likewise short-lived: for an energy
release $\Delta E\sim10^{-2}\Lambda$,
Eq.~\eqref{eq:baryon-cascade-lifetime} gives
$\tau_{\mB^\ast}=O(10^{-8}\,{\rm s})$, even before summing over multiple
final-state channels.  Thus all unstable $SU(\mN)$ hadrons decay well before
BBN, while $U(1)_{\mB}$ stabilizes the lightest neutral baryon.  

Assuming the meson-dominated late-stage 
completion of hadronization via 
string breaking discussed in 
Sec.~\ref{sec:hadronization}, negligible entropy injection, and
production-limited post-confinement evolution, the $\mN=12$ entry of
Table~\ref{tab:dm-matching} reproduces the observed DM density.  
The larger-$\mN$ entries describe complementary regimes.  For $\mN=15$ baryons can saturate the DM density for $m_{\mB}\simeq2.9\times10^{10}\GeV$, 
that is well beyond the  $\mathcal{O}(100)\,$TeV  
 bound implied by  unitarity of the DM annihilation cross section~\cite{Griest:1989wd,vonHarling:2014kha}.
For $\mN=18$ and $21$, the baryon mass required to saturate $\Omega_{\rm DM}$ corresponds to a confinement scale above the inflationary energy scale, so that any pre-existing  $\mB$ population  gets diluted away by inflation. Lowering $\Lambda$ correspondingly lowers $m_\mB$, implying that baryons can constitute only a tiny subcomponent of the DM.
These last cases can be relevant for post-inflation axion models: axions account for the entirety of DM, while large $\mN$ plays a double role, protecting the quality of the Peccei-Quinn symmetry~\cite{DiLuzio:2017tjx} and suppressing the abundance of cosmologically dangerous stable relics~\cite{DiLuzio:2026vhj}.
 Going beyond $\mN=21$ further suppresses  the
 relic density, but offers no new low-energy phenomenological implications.

\section{Detection prospects} 
\label{sec:detection}

The stable baryon is a compact composite state, with characteristic radius
$R_{\mB}\sim\Lambda^{-1}$. 
Since its constituents carry hypercharge, a neutral baryon can nevertheless
interact with photons through higher dimensional operators 
involving electromagnetic form factors, as extensively analyzed for electromagnetic and dipolar DM~\cite{Bagnasco:1993st,Banks:2010eh,Barger:2010gv,
Sigurdson:2004zp,Kavanagh:2018xeh,Ibarra:2024mpq,Kumar:2025wsp}.
The leading interactions with ordinary matter depend crucially on the baryon spin.  For even
$\mN$, the neutral ground state generally has  $J=0$ and hence the  
magnetic (and also electric) dipole moment vanish. Scattering can instead proceed via its charge
radius, electromagnetic polarizability, or the dimension-six portals in
Eq.~\eqref{eq:portals}.  
For odd $\mN$, the neutral spin-$1/2$ baryon can possess a magnetic dipole moment, corresponding to a dimension-five operator that can give rise to a parametrically less suppressed interaction. 
Magnetic moments of three-flavor composite states have been classified in Ref.~\cite{Aranda:2015jis}.

Detectability is also controlled by the relic baryon halo flux, which, due to the large-$\mN$  strong suppression   
of the baryon relic density, rapidly decreases 
with increasing $\mN$.
Consequently, for the largest values of $\mN$ the 
integrated flux of DM particles  might be too low to yield  any signal in  conventional  detectors.  
The $\mN=12$ benchmark is not flux limited and can be probed
through nuclear-recoil searches, whereas heavier states motivate dedicated
large-exposure and multiple-scatter searches for ultraheavy dark
matter~\cite{Kumar:2025wsp,Rajendran:2022kcs}.  At still smaller fluxes,
paleo-detectors, that exploit geological exposure times by searching for persistent damage tracks in ancient 
minerals~\cite{Baum:2021jak,Baum:2023cct}, or 
astrophysical 
probes~\cite{Bramante:2023djs,Phoroutan-Mehr:2025hjz}
can be used. 
However, observable multiple-scatter or paleo-detector signals, as well as compact-star heating~\cite{Bramante:2023djs} or 
the collapse of exoplanets into black holes~\cite{Phoroutan-Mehr:2025hjz}, all  require sufficiently large stopping power
and localized energy deposition.

\section{Interpretation and domain of validity}
\label{sec:validity}

The RRM provides a useful physical template for the two-body part of the
calculation.  A quark and antiquark form a resonant mesonic state respecting energy 
conservation~\cite{Ravagli:2007xx,Ravagli:2008rt,Fries:2025jfi}. 
Three-body extensions
use a diquark intermediate step to organize ordinary baryon formation~\cite{He:2019vgs}.  Our large-$\mN$ equations retain the two-body sequential approach as the leading formation chain, while neglecting simultaneous multi-body reactions, whose potential impact, especially at large $\mN$,  is not known.
Therefore  we cannot claim that a 12- or 15-step chain is
under  control,  merely because the one- and two-step QCD
applications are successful.
However, while multi-body reaction rates involving higher-order Casimir coefficients are expected to violate the simple quadratic Casimir scaling, there is no clear reason to expect them to avoid  bottleneck effects.
In any case,  this limitation does not erase the central physical hierarchy.  Any 
pairwise path starts with the $1/(\mN-1)$ amplitude ratio in
Eq.~\eqref{eq:first-hierarchy} and repeatedly competes with stronger meson-forming or cluster-destroying channels.  BNN fusion reduces the number
of steps required for baryon formation,
and provides an explicit test of 
multiquark-cluster shortcuts. 
The comparison of NN and BNN networks shows
 that the bottleneck is not an artifact of retaining only adjacent ranks.
Even for the largest values of $\mN$ studied, the BNN enhancement in $y_\mB$ over the NN result is only about one order of magnitude.
On the other hand, 
  the direct  projection of $\mN$ 
constituents of three different flavors 
onto the  spin--flavor symmetric baryon is always
$\binom{\mN+5}{5}/6^{\mN}$, independently of how that projection is factorized into binary steps.
This  ranges between $10^{-6}$ and $10^{-12}$ for $\mN =12$ and 21, respectively, providing an additional strong suppression.

There are further caveats to the interpretation of the results.  First, Eq.~\eqref{eq:rate-ansatz}
uses the squared quadratic Casimir with a channel-independent prefactor. Possible effects of 
higher invariants and representation-dependent kinematical coefficients are neglected.
Second, exact $SU(6)$ spin-flavor weights assume near-degenerate flavors and uniformly populated internal states.  Third, while the NN network is complete with respect to the attractive channels that form bound states, weakly attractive mixed-symmetry cluster channels that would appear in the BNN network have been omitted. Fourth, 
 the late string-breaking-dominated regime is not evolved dynamically, and the dominance of meson formation is justified only on the basis of qualitative arguments.
Within these approximations, all channel counting, symmetry factors, and numerical checks of conserved charges have been implemented exactly.

\section{Conclusions}

Cosmological evolution of populations of large-$\mN$ baryons need not begin with equilibrium densities.
Quadratic-Casimir scaling makes singlet meson formation and the destruction of small same-sign clusters parametrically stronger than the initial steps of baryon growth. Their repeated competition depletes the constituent reservoir, producing a Casimir bottleneck in the assembly of the $\mN$-quark singlet. The resulting strong suppression of the abundance of cosmologically stable baryons implies that, for sufficiently large $\mN$, baryon masses well above the $\mathcal{O}(100)\,\mathrm{TeV}$ unitarity limit on annihilation cross sections are required to saturate the DM density.

We have extended the NN description of this mechanism presented in the Letter~\cite{DiLuzio:2026aco} to all maximally attractive two-body fusions and rearrangements, which involve a tower of $L=0$ $\mN$-color antisymmetric clusters.
For $\mN> 3$ the additional reactions included in the  BNN network  enhance baryon formation by factors of $\mathcal{O}(10)$,  but do not remove the strong suppression identified with the NN network.
For $n_f=3$, the $SU(6)$ spin-flavor projection further strengthens the suppression. The BNN baryon pseudo-yield ranges from $y_\mB\sim 7.8\times10^{-14}$ for $\mN=12$ down to $y_\mB\sim 4.5\times10^{-31}$ for $\mN=21$.

The same three-flavor structure yields a realistic particle spectrum. For $\mN$ divisible by three and SM-like hypercharge assignments, all baryons have integer electric charge, and a neutral member exists. Hypercharge self-energy can make it the lightest state, which is cosmologically stable by virtue of an accidental $U(1)_\mB$ symmetry.
Anomaly couplings and dimension-six effective operators mediate the decays of mesons and charged baryons.
We have provided an $\mN=12$ spin-zero baryon benchmark in which, for an appropriately chosen mass, baryons account for all of the DM, while the unstable states decay well before BBN. 
For reasonable confinement scales, larger groups naturally yield stable relics that give a negligible contribution to the cosmological energy density. These scenarios provide useful frameworks for constructing high-quality axion models.

\section*{Acknowledgments}
We thank Marco Nardecchia for valuable contributions and constructive discussions
throughout the development of this project, 
and Giovanni Villadoro for useful conversations. We acknowledge the use of ChatGPT in coding the BNN network. The results have been systematically checked against independently written human code in the NN limit.
The work of LDL is supported by the Italian Ministry of University and Research (MUR) via the FIS2 Consolidator Grant project FIS-2023-02106 -- QAXION (CUP: I53C25001880001).
The work of EN is supported  by the Estonian Research Council grant PRG1884
and by the INFN ``Iniziativa Specifica'' Theoretical Astroparticle Physics (TAsP).
Partial support from  the Estonian Research Council grants TARISTU24-TK10, TARISTU24-TK3,  CoE grant TK202 “Foundations of the Universe”,
and from the CERN and ESA Science Consortium of Estonia, grants RVTT3 and RVTT7 is also acknowledged.


\appendix

\section{Casimir coefficients for the mixed-symmetry channels}
\label{app:group}

We start recalling the color factors for the 
completely antisymmetric channels used in the networks. 
Next,  we give  the Casimir coefficients for 
mixed-symmetry channels, that are not included in the networks. 

From Eq.~\eqref{eq:antisymmetric-Casimir}, projection of
$Q_p\otimes Q_q$ onto $Q_{p+q}$ gives
\begin{align}
 \mC^F_{pq}
 &=\frac12(C_{p+q}-C_p-C_q)\nonumber\\
 &=-\frac{\mN+1}{2\mN}pq,
 \qquad p+q\leq\mN.
 \label{eq:app-fusion}
\end{align}
For $p\geq q$, annihilating $q$ 
quarks against $q$ antiquarks leaves
$Q_{p-q}$, so
\begin{align}
 \mC^R_{p\bar q}
 &=\frac12(C_{p-q}-C_p-C_q)\nonumber\\
 &=-\frac{\mN+1}{2\mN}(\mN-p)q, \qquad q\leq p.
 \label{eq:app-rearrangement}
\end{align}

\subsection{Mixed-symmetry channles: one-quark reactions} 

The quadratic Casimir for the mixed-symmetry channel 
$Q_1 \otimes Q_p \to \mR_{p+1}$ with  Young tableu  
 $\mR_{p+1}\sim$ {\tiny $\substack{\yng(2,1)  \\ :\quad \\ \yng(1)\quad } $} \ is
\begin{eqnarray}
\nonumber
    C_{(p+1)_{\rm mix}} &=& \frac{p(\mN+1-p) +\mN+2}{2} -\frac{(p+1)^2}{2 \mN}\\
    &=& 1  -\frac{p}{\mN}+C_{p} +C_F\,.
    \quad
\end{eqnarray}
The second expression is convenient because 
it immediately  gives the Casimir coefficient: 
\begin{eqnarray}
    \mC_{p1}^{\rm mix} &=& 
    \frac{1}{2} \left( C_{(p+1)_{\rm mix}}  -  C_p-C_F\right) 
    = + \frac{\mN-p}{2\mN}\,.\qquad 
    \label{eq:qqp}
\end{eqnarray}
The process $\bar Q_1 + Q_p$ has  two channels,  
$\bar Q_1 \otimes Q_p \to Q_{p-1} + \mM$ which is 
maximally attractive, and the mixed-symmetry channel
$\bar Q_1 \otimes Q_p \to \mR_{p+\bar 1}$
 with Dynkin label $(0,0\dots,1_p,0\dots,1)$, where the subscript $p$ 
 labels the position of the entry. Its quadratic  Casimir is
\begin{eqnarray}
\nonumber
C_{(p+\bar 1)_{\rm mix}} &=& 
\frac{(\mN+1)(p(\mN-p)+\mN-1)+2p}{2\mN}\\
  &=& \frac{p}{\mN} + C_p + C_F \,.
\end{eqnarray}
This gives
 \begin{align}
     \mC_{p\bar 1}^{\rm mix} &= \frac{1}{2}\left(  
     C_{(p+\bar 1)_{\rm mix}}  - C_p - C_F \right) 
\label{eq:bqqp}
              = + \frac{p}{2\mN} \,.
\end{align}
\eq{eq:qqp} and \eq{eq:bqqp} show that both Casimir coefficients for the mixed-channels are positive. 
Therefore, in the NN network, all the 
channels not included are repulsive and thus, with respect to the attractive interactions
relevant to bound-state formation, the NN network is complete.

\subsection{Mixed-symmetry channels: cluster reactions}

The process $Q_p + Q_r \to \mR_{p+r}$
($p+r \leq \mN$)
has various channels, corresponding to the 
representation in the direct product
of the two antisymmetric representations 
$Q_p \otimes Q_r $.
Fusion into the antisymmetric $Q_{p+r}$ is 
the most strongly  attractive channel. The representations for 
the mixed states can be written as
\begin{equation}
    \mR_k = [2^k,1^{p+r-2k}],\quad k=0,1,\dots {\rm min}[p,r]\,.
\end{equation}
The Young tableau has $k$ rows of length 2 and $p+r-2k$ rows of length 1.
$k=0$ corresponds to a single column of $p+r$ boxes, that is the fully antisymmetric representation $Q_{p+r}$. 
The generic Casimir can be written as 
\begin{equation}
C^{\mR_k}_{p+r}=C_{p+r} + k(p+r+1-k)    \,.
\end{equation}
The  Casimir coefficient for the corresponding channel is
\begin{eqnarray}
\mC^{\mR_k}_{p+r}
\label{eq:qprk}
\!&=&\! \mC_{p+r} +\frac{k}{2}(p+r+1-k).
\end{eqnarray}
\eq{eq:qprk} shows that the $k=0$ 
antisymmetric channel is always the most attractive, although  certain  mixed-symmetry  channels  can also be  attractive.  

Let us now consider the process 
$Q_p + \bar Q_r \to \mR_{p-k +\bar r-\bar k} + k \mM$
($k\leq {\rm min}(p,r)$), with  $k$ color-contractions into mesons. 
Recalling that  $C_{\bar r} =C_{\mN-r}$, 
for $p-k \leq \mN - (r-k)$ (or equivalently 
$p+r \leq \mN+2k$) we obtain
%
\begin{eqnarray}
&&\hspace{-6mm} \mC^{\mR_k}_{p+\bar r}=
\frac{1}{2}\!\left[
\frac{(p-k)(r-k)}{\mN} +
C_{p-k} + C_{\mN-r+k} 
 - C_{p}  \nonumber 
  -C_{\mN-r}   
\right] \\
&&\  =
 + \frac{p\,r }{2\mN}-\frac{k}{2}(\mN+1 -p-r+k),  
\label{eq:qpbarrk1}
\end{eqnarray}
for $p+r\leq \mN+2k$.
Taking  for definiteness  $r\leq p$, this  can  be written as 
\begin{eqnarray}
    \mC_{\mR_k}^{p+\bar r} &=&  \mC^{R}_{p\bar r} 
+\frac{1}{2} \left[\mN+1 - (p-k)\right]\,(r-k), \qquad  
\end{eqnarray}
where  $\mC_{R}^{p\bar r} $,  given in \eq{eq:app-rearrangement}, 
is the  coefficient  of the most attractive channel, in which the maximum possible number $r$ of mesons are  formed. The second term 
on the RHS  is always positive. Therefore 
for the other mixed channels 
with $k<r$ the 
interaction is less attractive. Quark-antiquark recombination represents the attractive channel. For large values of $p$ and $r$ and small 
values of the recombination number $k$, the interaction governed by $ \mC_{\mR_k}^{p+\bar r}$ is repulsive.
For the other condition 
 $p-k \geq \mN - (r-k)$ (or equivalently 
$p+r \geq \mN+2k$) we obtain
\begin{eqnarray}
\hspace{-0pt}
\mC_{\mR_k}^{p+\bar r}\!&=&\!
+ \frac{(\mN-p)(\mN-r)}{2\mN}  \\
\nonumber
+ \frac{k}{2}[p\!\!&+&\!\!r-(\mN+k) +1]\,,
 \quad  (p+r\geq \mN+2k)\,,
\label{eq:qpbarrk2}
\end{eqnarray}
that is manifestly always positive. Therefore 
all the attractive channels lie in the branch 
$p+r\leq \mN+2k$.

\section{Network equations}
\label{app:network-checks}

Let us consider the NN network with $n_f=3$, 
and define  $s_p=s_6(1,p)$.  The  evolution equation 
for the density of free constituent is
\begin{equation}
 \dot y_1=y_1\left[
 |\hat\mC_{2,1}|^2y_2
 -\sum_{p=1}^{\mN-1}
 \left(|\hat\mC_{p,p-1}|^2
 +s_p|\hat\mC_{p,p+1}|^2\right)y_p
 \right].
 \label{eq:app-y1}
\end{equation}
For a $p$-cluster with $2\leq p\leq\mN$ we have
\begin{align}
 \dot y_p={}&y_1\Bigg[
 \frac{s_{p-1}}{1+\delta_{p2}}
 |\hat\mC_{p-1,p}|^2y_{p-1}
 +|\hat\mC_{p+1,p}|^2y_{p+1}
 \nonumber\\[-2mm]
 &\hspace{9mm}
 -\left(|\hat\mC_{p,p-1}|^2
 +s_p|\hat\mC_{p,p+1}|^2\right)y_p
 \Bigg],
 \label{eq:app-yp}
\end{align}
where the factor  $1/(1+\delta_{p2})$ avoids double counting in the 
$Q_1+Q_1\to Q_2$ process.
The following consistency conditions 
are imposed~\cite{DiLuzio:2026aco}
\begin{equation}
 \hat\mC_{\mN,\mN-1}
 =\hat\mC_{\mN,\mN+1}=0,
 \qquad y_{\mN+1}=0.
 \label{eq:app-boundary}
\end{equation}
For the BNN network, for  $\mN=3$ we have the eight reactions 
\begin{align}
 Q_1+Q_1&\to Q_2,&Q_1+Q_2&\to\mB,\nonumber\\
 \bar Q_1+\bar Q_1&\to\bar Q_2,
 &\bar Q_1+\bar Q_2&\to\bar\mB,\nonumber\\
 Q_1+\bar Q_1&\to\mM,
 &Q_1+\bar Q_2&\to\bar Q_1+\mM,\nonumber\\
 Q_2+\bar Q_1&\to Q_1+\mM,
 &Q_2+\bar Q_2&\to2\mM.
 \label{eq:su3-reactions}
\end{align}
Only the last reaction is not included 
in the  NN system. This  explains 
why for $\mN=3$ the downward BNN correction 
is small.

\section{The pNGBs and the decay channels}
\label{app:decays}

For $m_{U,D,S}\ll\Lambda$, the approximate chiral symmetry
$U(3)_L\times U(3)_R$ breaks to $U(3)_V$.  The octet consists of pNGBs,
while the axial singlet $\eta'_D$ also receives a mass from the hidden
topological susceptibility.  In the  basis $(U,D,S)$, define
\begin{align}
 t_3&=\frac12\operatorname{diag}(0,1,-1),
 &t_8&=\frac{1}{2\sqrt3}\operatorname{diag}(-2,1,1),
 \nonumber\\[-1mm]
 t_0&=\frac{\boldsymbol1}{\sqrt6}.
 \label{eq:generators}
\end{align}
With $\Tr(t_at_b)=\delta_{ab}/2$ and the hypercharge assignments $Y=\operatorname{diag}(2/3,-1/3,-1/3)$ 
 the mixed hypercharge anomalies are
\begin{align}
 \mathcal A_A&=2\mN\Tr(t_A Y^2),
 &\mathcal A_3&=0,
 \nonumber\\[-1mm]
 \mathcal A_8&=-\frac{2\mN}{3\sqrt3},
 &\mathcal A_0&=\frac{4\mN}{3\sqrt6}.
 \label{eq:anomalies}
\end{align}
The trace is over the  three-dimensional flavor space on which both the
chiral generators and the hypercharge matrix act.  $SU(\mN)$  color gives the
overall multiplicity $\mN$.
The anomalous octet width is, in the conventions of
Eq.~\eqref{eq:generators},
\begin{equation}
 \Gamma(\pi_8^0\to BB)
 =\frac{\mN^2\alpha_Y^2}{432\pi^3}
 \frac{m_{\pi_8}^3}{f_\pi^2},
 \label{eq:pi8-width}
\end{equation}
where $BB$ denotes a pair of $U(1)_Y$ gauge bosons. 
For a pNGB current normalized by
$\langle0|\bar\Psi_i\gamma^\mu\gamma_5\Psi_j|P(p)\rangle=if_Pp^\mu$,
a current-current operator in Eq.~\eqref{eq:portals} gives parametrically
\begin{equation}
 \Gamma(P\to f\bar f')\simeq
 \frac{N_c^f|c|^2}{8\pi}
 \frac{f_P^2m_P}{\Lambda_6^4}(m_f^2+m_{f'}^2),
 \label{eq:direct-pngb-width}
\end{equation}
where $c=\{a_U,a_{ij},b_i\}$, $f,f'$ are SM fermions, and possible 
phase space factors have been neglected. 
\eq{eq:direct-pngb-width} shows the expected chiral suppression.  Given the high $\Psi$ masses, top or heavy-flavor channels are open, and the suppression is not
cosmologically problematic.
The Witten--Veneziano relation gives
\begin{equation}
 m_{\eta'_D}^2=m_{\eta'_D,\,m}^2
 +\frac{6\chi_{\rm YM}}{f_{\eta'_D}^2},
 \label{eq:wv}
\end{equation}
with $m_{\eta'_D}^2/\Lambda^2=O(1/\mN)$ in the chiral large-$\mN$ limit
~\cite{Witten:1979vv,Veneziano:1979ec}.  Nevertheless, $\eta'_D$ is not
stable; it decays via the hypercharge anomaly:
\begin{equation}
 \Gamma(\eta'_D\to BB)
 \simeq \frac{\mN^2\alpha_Y^2}{216\pi^3}
 \frac{m_{\eta'_D}^3}{f_{\eta'_D}^2}.
 \label{eq:eta-width}
\end{equation}
With
$f_{\eta'_D}\sim\sqrt{\mN}\Lambda/(4\pi)$, the anomaly factors compensate
for the large-$\mN$ decrease of $m_{\eta'_D}$, giving parametrically
$\Gamma_{\eta'_D}/m_{\eta'_D}=O(\alpha_Y^2/\pi)$ in the anomaly-dominated
limit.

In the secluded $n_f=1$  model of~\cite{Morrison:2020yeg}, 
a long-lived axial-singlet population can
source baryons after confinement through the $2\to2$ freeze-in reaction $\eta'_D\eta'_D\to\mB\bar\mB$.  Here, 
 decays via the anomaly promptly remove  $\eta'_D$ from the thermal bath.  In addition, $m_{\mB}/m_{\eta'_D}=O(\mN^{3/2})$, so the reaction is strongly Boltzmann  suppressed for large $\mN$. Therefore  this secondary
baryon source can be safely neglected.

\bibliography{bibliography_merged}

\end{document}